\documentstyle[11pt]{article}
\newcounter{subequation}[equation]

\makeatletter 
 
\def\thesubequation{\theequation\@alph\c@subequation}
\def\@subeqnnum{{\rm (\thesubequation)}}
\def\slabel#1{\@bsphack\if@filesw {\let\thepage\relax
   \xdef\@gtempa{\write\@auxout{\string
      \newlabel{#1}{{\thesubequation}{\thepage}}}}}\@gtempa
   \if@nobreak \ifvmode\nobreak\fi\fi\fi\@esphack}
\def\subeqnarray{\stepcounter{equation}
\let\@currentlabel=\theequation\global\c@subequation\@ne
\global\@eqnswtrue
\global\@eqcnt\z@\tabskip\@centering\let\\=\@subeqncr
$$\halign to \displaywidth\bgroup\@eqnsel\hskip\@centering
  $\displaystyle\tabskip\z@{##}$&\global\@eqcnt\@ne
  \hskip 2\arraycolsep \hfil${##}$\hfil
  &\global\@eqcnt\tw@ \hskip 2\arraycolsep
  $\displaystyle\tabskip\z@{##}$\hfil
   \tabskip\@centering&\llap{##}\tabskip\z@\cr}
\def\endsubeqnarray{\@@subeqncr\egroup
                     $$\global\@ignoretrue}
\def\@subeqncr{{\ifnum0=`}\fi\@ifstar{\global\@eqpen\@M
    \@ysubeqncr}{\global\@eqpen\interdisplaylinepenalty \@ysubeqncr}}
\def\@ysubeqncr{\@ifnextchar [{\@xsubeqncr}{\@xsubeqncr[\z@]}}
\def\@xsubeqncr[#1]{\ifnum0=`{\fi}\@@subeqncr
   \noalign{\penalty\@eqpen\vskip\jot\vskip #1\relax}}
\def\@@subeqncr{\let\@tempa\relax
    \ifcase\@eqcnt \def\@tempa{& & &}\or \def\@tempa{& &}
      \else \def\@tempa{&}\fi
     \@tempa \if@eqnsw\@subeqnnum\refstepcounter{subequation}\fi
     \global\@eqnswtrue\global\@eqcnt\z@\cr}
\let\@ssubeqncr=\@subeqncr
\@namedef{subeqnarray*}{\def\@subeqncr{\nonumber\@ssubeqncr}\subeqnarray}
\@namedef{endsubeqnarray*}{\global\advance\c@equation\m@ne%
                           \nonumber\endsubeqnarray}

\makeatletter
\@addtoreset{equation}{section}
\makeatother
\renewcommand{\theequation}{\thesection.\arabic{equation}}

\def\dalemb#1#2{{\vbox{\hrule height .#2pt
        \hbox{\vrule width.#2pt height#1pt \kern#1pt
                \vrule width.#2pt}
        \hrule height.#2pt}}}

   \let\d=\delta 
  \let\q=\theta  
  \let\n=\nu   
\let\s=\sigma   \let\f=\phi \let\c=\chi 
\let\w=\omega      \let\G=\Gamma

\let\la=\label  
 \let\sse=\subsection  
\def\nn{\nonumber} \def\bd{\begin{document}} \def\ed{\end{document}}
\def\ds{\documentstyle} \let\fr=\frac \let\bl=\bigl \let\br=\bigr
\let\Br=\Bigr \let\Bl=\Bigl 
\let\bm=\bibitem
\let\na=\nabla
\let\pa=\partial \let\ov=\overline
\def\ie{{\it i.e.\ }} 
\newcommand{\be}{\begin{equation}} 
\newcommand{\ee}{\end{equation}} 
\def\ba{\begin{array}}
\def\ea{\end{array}}
\def\ft#1#2{{\textstyle{{\scriptstyle #1}\over {\scriptstyle #2}}}}
\def\fft#1#2{{#1 \over #2}}
\def\del{\partial}
\def\sst#1{{\scriptscriptstyle #1}}
\def\oneone{\rlap 1\mkern4mu{\rm l}}
\def\e7{E_{7(+7)}}
\def\td{\tilde}
\def\wtd{\widetilde}
\def\im{{\rm i}}
\def\bog{Bogomol'nyi\ }
\def\q{{\tilde q}}
\def\hast{{\hat\ast}}
\def\0{{\sst{(0)}}}
\def\1{{\sst{(1)}}}
\def\2{{\sst{(2)}}}
\def\3{{\sst{(3)}}}
\def\4{{\sst{(4)}}}
\def\5{{\sst{(5)}}}
\def\6{{\sst{(6)}}}
\def\7{{\sst{(7)}}}
\def\8{{\sst{(8)}}}
\def\n{{\sst{(n)}}}
\def\oo{{\"o}}
\def\hA{\hat{\cal A}}
\def\ns{{\sst {\rm NS}}}
\def\rr{{\sst {\rm RR}}}
\def\tH{{\widetilde H}}
\def\tB{{\widetilde B}}
\def\cA{{\cal A}}
\def\cF{{\cal F}}
\def\tF{{\wtd F}}
\def\v{{{\cal V}}}
\def\Z{\rlap{\sf Z}\mkern3mu{\sf Z}}
\def\ep{{\epsilon}}
\def\IIA{{\rm IIA}}
\def\IIB{{\rm IIB}}
\def\ads{{\rm AdS}}
\def\R{\rlap{\rm I}\mkern3mu{\rm R}}
\def\vp{{\varphi}}
\def\ns{{\sst{\rm NS}}}
\def\rr{{\sst{\rm RR}}}
 \def\cF{{\cal F}} \def\cA{{\cal A}} \def\cB{{\cal
B}} \def\hA{{\hat{\cal A}}} \def\td{\tilde} \def\wtd{\widetilde}
\def\e{{\epsilon}} 
\def\Z{\rlap{\sf Z}\mkern3mu{\sf Z}}

\def\hhs{{\qquad}}
\def\hs{\,\,}

\newcommand{\ho}[1]{$\, ^{#1}$}
\newcommand{\hoch}[1]{$\, ^{#1}$}
\newcommand{\bea}{\begin{eqnarray}} 
\newcommand{\eea}{\end{eqnarray}} 
\newcommand{\ra}{\rightarrow}
\newcommand{\lra}{\longrightarrow}
\newcommand{\Lra}{\Leftrightarrow}
\newcommand{\ap}{\alpha^\prime}
\newcommand{\bp}{\tilde \beta^\prime}
\newcommand{\tr}{{\rm tr} }
\newcommand{\Tr}{{\rm Tr} } 
\newcommand{\NP}{Nucl. Phys. }
\newcommand{\tamphys}{\it Center for Theoretical Physics,
Texas A\&M University, College Station, TX 77843}
\newcommand{\upenn}{\it Dept. of Physics and Astronomy, 
University of Pennsylvania,
Philadelphia, PA 19104}

\newcommand{\auth}{ C.N. Pope, A. Sadrzadeh and S.R. Scuro }

\begin{document}
\begin{flushright}
\hfill{CTP TAMU-21/99 \\ 
May 1999}\\
\hfill{\bf hep-th/9905161}\\
\end{flushright}


\begin{center}
{\large {\bf Timelike Hopf Duality and Type IIA$^*$ String Solutions} 
} 

\vspace{20pt}

\auth

\vspace{10pt}
{\hoch{}\tamphys}

\vspace{30pt}

\underline{ABSTRACT}
\end{center}

   The usual T-duality that relates the type IIA and IIB theories
compactified on circles of inversely-related radii does not operate if
the dimensional reduction is performed on the time direction rather
than a spatial one.  This observation led to the recent proposal that
there might exist two further ten-dimensional theories, namely type
IIA$^*$ and type IIB$^*$, related to type IIB and type IIA respectively
by a timelike dimensional reduction.  In this paper we explore
such dimensional reductions in cases where time is the coordinate of a
non-trivial $U(1)$ fibre bundle.  We focus in particular on situations
where there is an odd-dimensional anti-de Sitter spacetime
AdS$_{2n+1}$, which can be described as a $U(1)$ bundle over 
$\widetilde{CP}^n$, a non-compact version of $CP^n$ corresponding to
the coset manifold $SU(n,1)/U(n)$.  In particular, we study the
AdS$_5\times S^5$ and AdS$_7\times S^4$ solutions of type IIB supergravity 
and eleven-dimensional supergravity.  Applying a timelike Hopf
T-duality transformation to the former provides a new solution of the
type IIA$^*$ theory, of the form $\widetilde{CP}^2\times S^1\times S^5$.
We show how the Hopf-reduced solutions provide further examples of
``supersymmetry without supersymmetry.''  We also present a
detailed discussion of the geometrical structure of the Hopf-fibred 
metric on AdS$_{2n+1}$, and its relation to the horospherical metric
that arises in the AdS/CFT correspondence.

{\vfill\leftline{}\vfill
\vskip 10pt \footnoterule
{\footnotesize
        \hoch{}        Research supported in part by DOE 
grant DOE-FG03-95ER40917 \vskip -12pt}  \vskip  14pt
}

\pagebreak
\setcounter{page}{1}

\tableofcontents
\addtocontents{toc}{\protect\setcounter{tocdepth}{2}}
\newpage

\section{Introduction}

   The duality symmetries of string theory and M-theory have been
studied from a variety of viewpoints, in order to elucidate the
interconnections between theories once thought to be distinct.  The
duality relations can be either at the perturbative level, as in the
case of the T-duality that relates the type IIA and IIB string
theories compactified on circles of inversely-related radius
\cite{buscher,lag}, or non-perturbative, as in the strong-coupling
limit of the type IIA string leading to the ``opening out'' of an
eleventh dimension \cite{ht,wit}.

   In the case of a non-perturbative duality, the relation is
necessarily non-perturbative, and strictly speaking can be regarded
only as conjectural at the present time.  Consequently, it is of
interest to accumulate a body of evidence that supports the duality
conjecture.  One of the most useful tools in this regard has been the
study of BPS-saturated solutions to the low-energy effective actions, which
may be interpreted as solitonic states in the non-perturbative
spectra of the theories (see, for example, \cite{dkl}). 

    One of the commonly-employed techniques for uncovering the web of
interrelations between the various theories is to study the spectrum
of solutions corresponding to branes ``wrapped around'' non-trivial
cycles in an internal compactifying manifold.  The simplest such
manifolds are tori.  In this procedure, otherwise known as diagonal
dimensional reduction, a $p$-brane in $D$ spacetime dimensions wrapped
around an $n$-dimensional torus reduces to a $(p-n)$-brane in $D-n$
dimensions.  Another possible reduction procedure involves first
preparing a periodic array of $p$-branes in $D$ dimensions, so that
from a large-distance standpoint the $p$-brane is effectively
``smeared'' uniformly over an $n$-dimensional submanifold of its
$(D-p-1)$-dimensional transverse space.  These dimensions effectively
correspond to Killing directions in the transverse space, and can then
be wrapped around the $n$-torus.  This procedure, in which the
$p$-brane retains its spatial dimension $p$, while the spacetime
dimension is again reduced from $D$ to $D-n$, is known as vertical
dimensional reduction.  One can also, of course, consider situations
where the reduction on the $n$-torus is a mixture of diagonal and
vertical dimensional reduction steps.

    Another kind of reduction on a circle is also possible, in certain
cases.  The space transverse to a $p$-brane can be described in terms
of hyperspherical polar coordinates, with a radial coordinate $r$ and
angular coordinates on the foliating spheres.  If the transverse space
is even-dimensional, these spheres will be odd-dimensional, and then
we can exploit the fact that the sphere $S^{2n+1}$ can be described as
a $U(1)$ bundle over $CP^n$.  The coordinate $\psi$ on the $U(1)$
fibre is a Killing direction, and the Killing vector $\del/\del\psi$
is of constant length in the sphere metric.  It thus provides a
suitable coordinate for an $S^1$ dimensional reduction.  This
reduction mechanism, known as Hopf reduction \cite{dlp1,dlp2,dlp3},
has been studied in a variety of contexts.  It is akin to a vertical
dimensional reduction, in that it is the dimension of the transverse
space that is reduced.  However, it is quite different from the usual
vertical reduction in that $\del/\del\psi$ is a Killing direction even
for a single isotropic $p$-brane.

   In certain cases, such as the D3-brane of the type IIB theory, an
important phenomenon is that as the horizon of the brane is
approached, the spacetime geometry more and more nearly approaches the
product of an anti-de Sitter (AdS) spacetime and a sphere.  Specifically,
the sphere corresponds to the foliating spheres of the transverse
space, while the AdS spacetime is described in terms of foliating
surfaces that are formed by the $p$-brane world-volume with the
original radial coordinate $r$ now parameterising this sequence of
spacetime foliations.  In the case of the D3-brane, the near-horizon
geometry is therefore AdS$_5\times S^5$.  The description of AdS as a
foliation of flat Minkowski spacetime surfaces is known as the
horospherical construction of AdS.  The structure of this
near-horizon geometry plays a central r\^ole in the conjectured
AdS/CFT correspondence, where the degrees of freedom in the bulk
theory are mapped to those of a conformal field theory on the flat
boundary of AdS \cite{mald,gkp,wit2}. 

   In \cite{dlp2}, the Hopf reduction on the $U(1)$ fibres of the $S^5$ in
this near-horizon geometry was investigated.  One obtains a solution
of the nine-dimensional theory with an AdS$_5\times CP^2$ geometry.
Redefining fields according to the map that relates the type IIB and
type IIA variables in $D=9$, one finds, upon lifting the solution back
to $D=10$, a solution of the type IIA theory with the geometry
AdS$_5\times S^1\times CP^2$.  In other words, the interchange of
Kaluza-Klein and winding modes under T-duality has resulted in an
``untwisting'' of the $U(1)$ fibres in $S^5$.  One of the intriguing
features of this process is that one arrives at a solution of the type
IIA theory that ostensibly has no supersymmetry, and indeed, no
fermions at all.  The reason for this is related to the fact that
$CP^2$ does not admit an ordinary spin structure.  Rather, it admits a
generalised spin structure \cite{hp}, meaning that all fermions must
necessarily carry certain specific non-zero values of electric charge
coupling to the $U(1)$ gauge field whose connection is responsible for
the twisting of the $U(1)$ bundle.  In other words, all the fermions
in the nine-dimensional theory must be charged with respect to the
Kaluza-Klein vector of the type IIB reduced theory.  After the
T-duality transformation, this becomes the winding vector, and so in
consequence, from the ten-dimensional type IIA point of view the
fermions would all have to carry non-zero winding-mode charge. 
They would all be associated with non-trivial winding states, and so
would be seen only at the level of the full string theory, rather than
in the low-energy effective supergravity \cite{dlp2}.  Thus the
AdS$_5\times S^1\times CP^2$ solution of type IIA supergravity is
actually a perfectly respectable BPS solution of the full string
theory, which happens to look rather barren and unsatisfactory from
the field-theory standpoint.  

    As is well known, the anti-de Sitter spacetimes AdS$_n$ are
closely analogous to the spheres $S^n$, with the only difference that
they are described by constant-radius surfaces in a flat embedding
space $\R^{n+1}$ with metric signature $(2,n-1)$ rather than
$(0,n+1)$.  By the same token, therefore, the anti-de Sitter
spacetimes AdS$_{2n+1}$ of odd dimension can be described in terms of
$U(1)$ fibrations over certain spaces $\wtd{CP}^n$, which are
non-compact versions of the usual complex projective spaces $CP^n$.
The fibre coordinate is actually the time coordinate of AdS$_{2n+1}$,
so the non-compact $\wtd{CP}^n$ is still a space of Euclidean metric
signature.\footnote{It is clear that the fibre must be timelike in
this construction, since the metric of any complex space such as
$\wtd{CP}^n$ must necessarily have an {\it even} number of timelike
directions.  Thus the single timelike direction of AdS$_{2n+1}$ 
resides in the fibre, and not in the complex base-manifold.}  It
should be emphasised that this time coordinate on the fibre is
periodic, and so the bundle is $U(1)$ even in this construction of the
non-compact AdS spacetime.

    Using this observation, we can now follow a similar strategy to
the one used in \cite{dlp2}, but now we can untwist the fibres of the
AdS$_5$ in the AdS$_5\times S^5$ solution of the type IIB theory.
This will give a $\wtd{CP}^2\times S^5$ solution of the
Euclidean-signatured nine-dimensional theory obtained by time
reduction of the type IIB theory.  It is important, in this regard,
that the time coordinate on the Hopf fibres is naturally periodic, as
is appropriate for a Kaluza-Klein reduction accompanied by a
truncation to the ``massless'' sector. (See, for example, \cite{hj,cllpst}
for discussions of timelike reductions in supergravity.)  
As was shown in \cite{cllpst}, the
timelike reductions of the type IIA and type IIB theories in $D=9$ are
{\it not} related by (real) field redefinitions, and thus we cannot
perform such a redefinition in $D=9$ to obtain a solution of the usual
type IIA theory in $D=10$.  However, as was shown in
\cite{hull1,hull2}, one can
instead postulate the existence of a different theory in $D=10$, called type
IIA$^*$ in \cite{hull1}, which {\it is} related by timelike
T-duality to the type IIB theory.  The IIA$^*$ theory has a normal
$(1,9)$ metric signature, but its Ramond-Ramond fields have kinetic
terms of the ``wrong'' sign.   Similarly, a type IIB$^*$ theory was
also postulated in \cite{hull1}, related by timelike T-duality to the usual
type IIA theory.  

    The upshot of the above discussion is that if we lift the
nine-dimensional $\wtd{CP}^2\times S^5$ solution of the
nine-dimensional Euclidean-signatured theory back to $D=10$ after
performing the redefinition to IIA variables, we will obtain an
$S^1\times \wtd{ CP}^2\times S^5$ solution not of the type IIA theory,
but rather of the type IIA$^*$ theory postulated in \cite{hull1}.  The time
coordinate of this ten-dimensional solution lies along the $S^1$
direction.  Like the previously-discussed AdS$_5\times S^1\times CP^2$
solution of the usual type IIA theory, it is a solution that
ostensibly lacks not only supersymmetry but also fermions, since
$\wtd{CP}^2$ also does not admit an ordinary spin structure.  However,
again this is an artefact of looking only at the field-theoretic tip
of the string-theoretic iceberg, and in the full theory it should
correspond to an honest BPS configuration.

     In this paper, we explore some of the consequences of making Hopf
reductions on the timelike fibre coordinate of odd-dimensional AdS
spacetimes.  We begin in section 2 by setting up our notation and
using it to describe the ten-dimensional type IIA$^*$ theory of
\cite{hull1}, and its relation to type IIB.  In section 3, we describe
the geometry of the non-compact $\wtd{CP}^n$ spaces, including a very
simple coordinatisation that is somewhat analogous to the
horospherical coordinatisation of AdS itself.  We also show how the
odd-dimensional AdS$_{2n+1}$ metrics are described as $U(1)$
fibrations over $\wtd{CP}^n$.  We then discuss the geometrical
structure of the Hopf-fibred description of AdS$_{2n+1}$, and its
relation to the horospherical parameterisation. In section 4 we
discuss the supersymmetry of the AdS$_5\times S^5$ solution, and its
T-duality related $S^1\times \wtd{CP}^2\times S^5$ solution, making
use of results for the Killing spinors in AdS$_5$ that are derived in
an appendix.  In section 5, we discuss the AdS$_7\times S^4$ solution
of eleven-dimensional supergravity, and relate it to a
$\wtd{CP}^3\times S^4$ solution of the Euclidean-signatured
ten-dimensional supergravity coming from the timelike reduction of
$D=11$.  We discuss the supersymmetry of the ten-dimensional solution,
using results for the Killing spinors of AdS$_7$ that are also
obtained in the appendix.

\section{Type IIB/IIA$^*$ T-duality}

 In this section we consider the case of timelike reductions and
T-duality. The type IIB theory on a timelike circle of radius $R$
should be T-dual to some string theory on a timelike circle of
radius $1/R$, so that the limit $R \ra 0$ should give a T-dual
string theory in $9+1$ non-compact dimensions. This T-duality
cannot relate the IIB theory reduced on a timelike circle to the usual
type IIA theory reduced on the dual
timelike circle, since as was shown in \cite{cllpst}, the usual IIB and IIA
theories are not equivalent in $D=9$, after timelike reductions.
Instead, one is led to propose a new theory, called type IIA$^*$, in
$D=10$, whose timelike reduction is equivalent, after field
redefinitions, to the timelike reduction of the usual type IIB theory
\cite{hull2}.   We shall begin by setting up our notation and conventions.

    Let us begin with the type IIB theory in $D=10$.  The bosonic equations of
motion can be derived from the Lagrangian
\bea
{\cal L} &=& R\, {*\oneone} - \ft12 {*d\phi}\wedge d\phi - \ft12
e^{2\phi}\, {*d\chi}\wedge d\chi-\ft12 e^{-\phi}\,
{*G_\3^\ns}\wedge G_\3^\ns -\ft12
e^{\phi}\, {*G_\3^\rr}\wedge G_\3^\rr\nn\\
&&- \ft14 {*G_\5}\wedge G_\5   + \ft12 B_\4\wedge dB_\2^\ns \wedge
dB_\2^\rr
\ ,\label{2blag}
\eea
where the various field strengths are defined by
\bea
G_\3^\ns &=& dB_\2^\ns\ ,\qquad G_\3^\rr = dB_\2^\rr -\chi\, B_\2^\ns
\ ,\nn\\
G_\5 &=& dB_\4 + \ft12 B_\2^\ns\, dB_\2^\rr -\ft12 B_\2^\rr\, dB_\2^\ns 
\ .\label{2bfields}
\eea
As described in \cite{bbo}, the self-duality of $G_\5$ is to be
imposed here after varying the Lagrangian (\ref{2blag}) to obtain the
equations of motion.  This can be done consistently, since the
equation of motion for $G_\5$ turns out to be $d{*G_\5}= dB_\2^\ns\,
dB_\2^\rr$, and the right-hand side is
identical to the expression for the Bianchi identity for $G_\5$,
following from (\ref{2bfields}).

    Upon reduction on the time coordinate to $D=9$, this gives
\bea
{\cal L}_9 &=& R\, {*\oneone} - \ft12 {*d\phi}\wedge d\phi - \ft12
{*d\vp \wedge d\vp} - \ft12 e^{2\phi}\, {*d\chi}\wedge d\chi
+\ft12 e^{-8\alpha\vp}\, {*G_\4}\wedge G_\4 \nn\\
&&- \ft12 e^{-\phi+4\alpha\vp}\, {*G_\3^\ns}\wedge G_\3^\ns
+ \ft12 e^{-\phi-12\alpha\vp}\, {*G_\2^\ns}\wedge G_\2^\ns\nn\\
&& - \ft12 e^{\phi+4\alpha\vp}\, {*G_\3^\rr}\wedge G_\3^\rr
 + \ft12 e^{\phi-12\alpha\vp}\, {*G_\2^\rr}\wedge G_\2^\rr
+\ft12 e^{16\alpha\vp}\, {*\cF_\2}\wedge \cF_\2 \label{2bd9lag}\\
&&
+\ft12 B_\3\, dB_\2^\ns\, dB_\2^\rr - \ft12 B_\4\, dB_\1^\ns\,
dB_\2^\rr + \ft12 B_\4\, dB_\2^\ns\, dB_\1^\rr \nn\\
&& -\ft12 \chi\, B_\3\, dB_\2^\ns\, B_\2^\ns\ ,\nn
\eea
where
\bea
&&G_\1 = d\chi\ , \qquad G_\2^\rr = dB_\1^\rr - \chi\, dB_\1^\ns\ ,\nn\\
&&G_\2^\ns = dB_\1^\ns\ ,\qquad
G_\3^\ns = dB_\2^\ns - \cB_\1\, dB_\1^\ns\ ,\nn\\
&&G_\3^\rr = dB_\2^\rr -\chi\, dB_\2^\ns - \cB_\1\, dB_\1^\rr +
\chi\, \cB_\1\, dB_\1^\ns \ ,\nn\\
&&G_\4 =dB_\3 - \ft12 B_\1^\ns\, dB_\2^\rr + \ft12 B_\2^\ns\,
dB_\1^\rr + \ft12 B_\1^\rr\, dB_\2^\ns - \ft12 B_\2^\rr\, dB_\1^\ns
\ , \nn\\
&&G_\5 = dB_\4 + \ft12 B_\2^\ns\, dB_\2^\rr -\ft12 B_\2^\rr\,
dB_\2^\ns - \cB_\1\, dB_\3 -\ft12 B_\1^\ns\, \cB_\1\, dB_\2^\rr \nn\\
&&\qquad\quad
 - \ft12 B_\2^\ns\, \cB_\1\, dB_\1^\rr +
\ft12 B_\1^\rr\, \cB_\1\, dB_\2^\ns +\ft12 B_\2^\rr\, \cB_\1\,
dB_\1^\ns\ ,\nn\\
&& \cF_\2 = d\cB_\1\ .
\eea
Note that the kinetic terms for $G_\4$, $G_\2^\ns$, $G_\2^\rr$ and $\cF_\2$
all have signs that are reversed compared with the case of a standard spatial
reduction.  

    In a standard spatial reduction, the above nine-dimensional fields
would be equivalent, after appropriate redefinitions, to those coming
from the spatial reduction of the type IIA theory.  In Table 1 below,
we indicate how the various potentials should be related.  However, to
indicate the changes resulting from performing the reduction on the
time, direction, we underline those fields that arise with the
``non-standard'' sign for their kinetic terms.  Our notation
for the reduction of the gauge potentials $A_\3$, $A_\2$ and $A_\1$ of
the type IIA$^*$ theory to $D=9$ is that each $A_\n$ reduces to give
$A_\n$ and $A_{\sst{(n-1)}1}$.

\bigskip\bigskip
\begin{center}
\begin{tabular}{|c|c|c|c|c|c|}\hline
    &\multicolumn{2}{|c|}{IIA$^*$} &
    &\multicolumn{2}{c|}{IIB} \\ \cline{2-6}
    & $D=10$ & $D=9$ &T-duality & $D=9$ & $D=10$ \\ \hline\hline
    & $\underline{A}_\3$ & $\underline{A}_\3$ & $\longleftrightarrow$ &
                   $\underline{B}_\3$ & $B_\4$ \\ \cline{3-6}
R-R & &  $A_{\2 1}$& $\longleftrightarrow$
                           & $B_\2^{\rr}$ & $B_\2^{\rr}$
                                               \\ \cline{2-5}
fields& $\underline{A}_\1$ & $\underline{A}_\1$ &
                $\longleftrightarrow$ &
        $\underline{B}_\1^{\rr}$ & \\ \cline{3-6}
   & & ${A}_{\0 1}$ & $\longleftrightarrow$
                            & $\chi$ &$\chi$
                                 \\ \hline\hline
NS-NS & $g_{MN}$ & $\underline{\cal A}_\1$
                        & $\longleftrightarrow$ &
        $\underline{B}_\1^{\ns}$ & $B_\2^{\ns}$ \\ \cline{2-5}
fields& $A_{\2}$ & $A_{\2}$ &
               $\longleftrightarrow$ & $B_\2^{\ns}$ &
                                       \\ \cline{3-6}
      & & $\underline{A}_{\1 1}$ & $\longleftrightarrow$ &
                              $\underline{\cal B}_\1$ & $g_{MN}$
                                       \\ \hline
\end{tabular}
\end{center}
\bigskip\bigskip

\centerline{Table 1: Gauge potentials of type IIA$^*$ and IIB theories
             in $D=10$ and $D=9$}
\bigskip\bigskip

   The remarkable observation that allowed Hull to hypothesise the
existence of a new type IIA$^*$ string theory is that after making the
identifications of fields in $D=9$, as indicated in Table 1, it turns
out that the fields of the type IIA formulation in $D=9$ can be
reassembled into fields in $D=10$, where the fields $A_\3$ and $A_\1$
of the RR sector in $D=10$ have the non-standard sign for their
kinetic terms.\footnote{It was not {\it a priori} obvious that this
should have been possible, since it requires that the signs of the
kinetic terms of the nine-dimensional fields after transforming to
the type IIA language must be exactly correct in order to allow them
to reassemble covariantly as ten-dimensional fields.}
This theory, which is called the type IIA$^*$
theory, therefore has the low-energy effective Lagrangian
\bea
{\cal L} &=& R\, {*\oneone} -\ft12 {*d\phi}\wedge d\phi + \ft12
e^{\fft32\phi}\, {*F_\2}\wedge F_\2 - \ft12 e^{-\phi}\,
{*F_\3}\wedge F_\3 + \ft12 e^{\fft12\phi}\, {*F_\4}\wedge F_\4 \nn\\
&&\!\!
-\ft12 dA_\3\wedge dA_\3 \wedge A_\2 \ ,
\label{2alag}
\eea
It should be emphasised that the type IIA$^*$ theory in $D=10$ has a
normal $(1,9)$ spacetime signature.

    For completeness, we note that by the same token, one can also 
propose a type IIB$^*$ theory, which again differs from the usual type
IIB theory only in having the non-standard signs for the kinetic terms
of its RR fields \cite{hull1,hull2}.  It follows that the type IIA$^*$ and type
IIB$^*$ theories are themselves related by an ordinary spacelike
T-duality.

   Our interest will be in starting from a solution of the usual type
IIB theory, with the geometry AdS$_5 \times S^5$, and performing a
timelike reduction to $D=9$.  After redefining fields as in Table 1,
we can then lift the solution back to $D=10$ to obtain a solution of
the type IIA$^*$ theory.  In particular, our timelike reduction will
be of a slightly unusual kind, in which time is the fibre coordinate
on AdS$_5$ viewed as a Hopf bundle over $\wtd{CP}^2$, where
$\wtd{CP}^2$ is a certain non-compact form of the complex projective
2-space.  Such a reduction is a timelike analogue of the spacelike
Hopf reductions that were considered in \cite{dlp2}.  Having arrived at a
$\wtd{CP}^2 \times S^5$ solution in $D=9$, we may lift it back to $D=10$ 
to give a $\wtd{CP}^2 \times S^1 \times S^5$ solution of the type
IIA$^*$ theory.

    Let us begin with the AdS$_5\times S^5$ solution of the type
IIB theory. This solution involves just the metric tensor and the self-dual
5-form field strength $G_\5$ of the type IIB theory, whose relevant
equations of motion can be written simply as
\bea
R_{MN} &=& \ft1{96}\, G_{MPQRS}\, G_N{}^{PQRS}\ ,\nn\\
G_\5 &=& {*G_\5}\ .\label{gheq}
\eea
In the absence of the other fields of the theory, we have
simply $G_\5=dB_\4$.  We may find a solution on AdS$_5\times S^5$ of
the form
\bea
ds^2_{10} &=& ds^2_{AdS_5} + ds^2_{S^5}\ ,\nn\\
G_\5 &=& 4m \ep_{AdS_5} + 4m \ep_{S^5}\ ,\label{2bconfig}
\eea
where $\ep_{AdS_5}$ and $\ep_{S^5}$ are the volume forms for the
metrics $ds^2_{AdS_5}$ and $ds^2_{S^5}$ on AdS$_5$ and $S^5$
respectively, $m$ is a constant, and the metrics on AdS$_5$ and $S^5$
satisfy
\be
R_{\mu\nu} = -4 m^2\, g_{\mu\nu}\ ,\qquad
R_{mn} = 4 m^2\, g_{mn}
\ee
respectively, where the indices run from 1 to 5.  Since the unit AdS$_5$
has metric $d\Omega^2_{AdS_5}$ with Ricci tensor $\bar R_{mn} = - 4\,
\bar g_{mn}$, it follows that we can write
\be
ds^2_{AdS_5} = \ft1{m^2}\, d\Omega^2_{AdS_5}\ .
\ee
By Hopf fibering AdS$_5$ over $\wtd{CP}^2$, we can write the
line element as
\be
ds^2_{AdS_5} = \ft1{m^2}\, d{\wtd \Sigma}_4^2 - \ft1{4m^2}\,
(d\tau+\bar{\cal A})^2\ ,
\ee
where $d{\wtd \Sigma}_4^2$ is the metric on the ``unit'' $\wtd{CP}^2$,
and $d\bar{\cal A}= 4J$, where $J$ is the K\"ahler form on $\wtd{CP}^2$.

     We may now perform a timelike dimensional reduction of this
solution to $D=9$, along the $U(1)$ fiber parameterized by $\tau$.
Comparing with the general Kaluza-Klein prescription (with the dilaton
set to zero), for which
\bea
ds_{10}^2 &=& ds_9^2 - (dt +{\cal B}_\1)^2\ ,\nn\\
G_\5(x, t) &=& G_\5(x) + G_\4(x)\wedge (dt + {\cal B}_\1(x))\ ,
\eea
where $t=\ft1{2m} \tau$ and ${\cal B}_\1=\ft1{2m} {\bar {\cal A}}$,
we see, from the fact that the volume forms of the unit 
AdS$_5$ and $\wtd{CP}^2$ are
related by $\Omega_{AdS_5} = \ft12(d\tau+\bar{\cal A})\wedge {\wtd \Sigma}_4$,
that the solution will take the 9-dimensional form
\bea
ds_9^2 &=& ds^2_{S^5} + \ft1{m^2}\, d{\wtd \Sigma}_4^2\ ,\nn\\
F_\4 &=& \ft{4}{m^3}\, {\wtd \Sigma}_4\ ,\qquad {\cal F}_\2 = \ft2{m}\, J\ .
\label{d92bsol}
\eea
We note that in the dimensional reduction of the 5-form of the type IIB
theory, its self-duality translates into the statement that the fields
$G_\5$ and $H_\4$ in $D=9$ must satisfy $G_\4 ={* G_\5} = F_\4$.

     We now perform the T-duality transformation to the fields of the
$D=9$ reduction of the type IIA$^*$ theory.  The relation between the IIB
and the IIA$^*$ fields is shown in Table 1.
Thus in the IIA$^*$ notation, we have the nine-dimensional configuration
\bea
ds_9^2 &=& ds^2_{S^5} + \ft1{m^2}\, d{\wtd \Sigma}_4^2\ ,\nn\\
F_\4 &=& \ft4{m^3}\, {\wtd \Sigma}_4\ ,\qquad F_{\2 1} = \ft2{m}\, J\ .
\label{d92bsol2}
\eea
The crucial point is that the 2-form field strength $F_{\2 1}$ of
the IIA$^*$ variables is no longer a Kaluza-Klein field coming from the
metric; rather, it comes from the dimensional reduction of the 3-form
field strength in $D=10$.  Indeed, if we lift the solution
(\ref{d92bsol2}) back to $D=10$, we have the type IIA$^*$ configuration
\bea
ds_{10}^2 &=& ds^2_{S^5} + \ft1{m^2}\, d{\wtd \Sigma}_4^2 - dt^2\ ,\nn\\
F_\4 &=& \ft4{m^3}\, {\wtd \Sigma}_4\ ,\qquad F_{\3 1} = \ft2{m}\, J
\wedge dt\ .\label{2ad10}
\eea
The solution has the topology ${\wtd{CP}^2}\times S^1\times S^5$.  This
should be contrasted with the topology AdS$_5\times S^5$ for the
original $D=10$ solution in the type IIB framework.  Thus the
T-duality transformation in $D=9$ has ``unravelled'' the twisting of
the $U(1)$ fiber bundle over $\wtd{CP}^2$, leaving us with a direct product
${\wtd{CP}^2}\times S^1$ in the type IIA$^*$ description.

   In order to study the geometry of this solution in more detail, we
shall now present an explicit construction of the $\wtd{CP}^2$, and
more generally, $\wtd{CP}^n$, metrics.

\section{$\wtd{CP}^n$ and the Hopf fibration of AdS$_{2n+1}$}

\subsection{Fubini-Study metric on $\wtd{CP}^n$, and the coset
$SU(n,1)/U(n)$}

    We shall begin by constructing the metric on AdS$_{2n+1}$,
written as a $U(1)$ bundle over $\wtd {CP}^n$.  This will
closely parallel the standard construction of the Fubini-Study metric
on the usual compact $CP^n$.  

    To begin, we introduce $(n+1)$ complex coordinates ${Z}^a$ on
$C^{\, n+1}$, where $0\le a \le n$.     We also introduce the metric
$\eta_{ab}$, defined by
\be
\eta_{ab}= {\rm diag}\, (-1, 1, 1,\ldots, 1)\ .\label{eta}
\ee
The line element on $C^{\, n+1}$ may be taken to be:
\be
d\hat s^2 = \eta_{ab}\, d{Z}^a\, d\bar {Z}^b\ ,
\label{flat}
\ee
and is obviously invariant under $SU(n,1)$.  We then impose the
$SU(n,1)$-invariant constraint 
\be
\eta_{ab}\, {Z}^a\, \bar {Z}^b = -1\ ,\label{const}
\ee
which restricts the metric to the AdS$_{2n+1}$ manifold. 

   Splitting the indices $a=(\ell,0)$, with $1\le \ell\le n$, we now
introduce inhomogeneous complex coordinates $\zeta^\ell$,
defined by
\be
\zeta^\ell = {Z}^\ell/{Z}^0\ ,\qquad 1\le \ell\le n\ .\label{inhomo}
\ee
The line element  (\ref{flat}), subject to the constraint (\ref{const}),
can then be expressed in terms of the coordinates $\zeta^\ell$ and ${
Z}^0$ as follows:
\be
d\hat s^2 = -|d{Z}^0|^2 + |{Z}^0|^2\, d\zeta^\ell\,
d\bar\zeta^\ell + |\zeta|^2\, |d{Z}^0|^2 + {Z}^0\, \bar\zeta^\ell\,
d\zeta^\ell\, d\bar {Z}^0 + \bar {Z}^0\,\zeta^\ell\, d\bar\zeta^\ell\,
d{Z}^0\ , \label{xxx}
\ee
where $|\zeta|^2$ denotes $\zeta^\ell\, \bar\zeta^\ell$.
One can rewrite (\ref{xxx}) by completing the square, as
\be
d\hat s^2 = - \Big| \fft{d{Z}^0}{{Z}^0} -
|{Z}^0|^2\, \bar\zeta^\ell\, d\zeta^\ell \Big|^2 + |{Z}^0|^2\,
d\zeta^\ell\, d\bar\zeta^\ell + |{  Z}^0|^4\, \bar\zeta^\ell\, \zeta^m\,
d\zeta^\ell\, d\bar \zeta^m\ . \label{s2n1}
\ee
This is just like a Kaluza-Klein metric, with
the first term corresponding to the extra dimension.  We may think of the
$U(1)$ coordinate as being $\tau$, where ${  Z}^0=|{  Z}^0|\,
e^{\fft{\im}{2}\tau}$. Thus we can read off the metric on the base space,
orthogonal to $\del/\del\tau$, by dropping
the first term in (\ref{s2n1}), giving us the line element
\be
ds^2_{\wtd CP^n} = |{  Z}^0|^2\, d\zeta^i\, d\bar\zeta^i +
|{  Z}^0|^4\, \bar\zeta^i\, \zeta^j\, d\zeta^i\, d\bar \zeta^j\ .
\label{almost}
\ee
Noting from (\ref{const}) and (\ref{inhomo}) that we have
$|{  Z}^0|^2 = (1 - |\zeta|^2)^{-1}$, 
we see that the metric on the base space is
\be
ds^2_{\wtd CP^n} = \fft{d\zeta^\ell\, d\bar\zeta^\ell}{1 - |\zeta|^2} +
  \fft{ \bar\zeta^\ell\,\zeta^m\, d\zeta^\ell\, d\bar \zeta^m}{
 (1 - |\zeta|^2)^2}\ .\label{fubstud}
\ee
This is the metric on $\widetilde{CP}^n$, and is analogous to the
standard Fubini-Study metric on the standard compact complex
projective space $CP^n$.

    A convenient parameterisation for $\wtd{CP}^n$ can be obtained by
introducing the $2n+1$ real coordinates $(\tau, \phi,\chi, x_i,y_i)$, and
writing the homogeneous coordinates of $\wtd{CP}^n$ as
\bea
Z^0 &=& e^{\fft{\im}2\tau}\, 
\Big(\cosh\ft12\phi + \ft18 e^{\fft12\phi}\, (4\im\, \chi - 2\im \,
x_i\, y_i\, + x_i^2 + y_i^2)\Big)\ ,\nn\\
Z^n &=& e^{\fft{\im}2 \tau}\, 
\Big( \sinh\ft12\phi - \ft18 e^{\fft12\phi}\, (4\im\, \chi - 2\im \,
x_i\, y_i\, + x_i^2 + y_i^2) \Big)\ ,\label{xycoord}\\
Z^i &=& \ft12 e^{\fft12(\phi+\im\, \tau)}\, 
(x_i + \im\, y_i)\ , \quad 1\le i\le
n-1 \ .\nn
\eea
It is easily verified that these $Z^a$ satisfy the constraint
(\ref{const}).  Substituting into (\ref{flat}), we immediately obtain
the metric on AdS$_{2n+1}$, in the form
\be
d\Omega^2_{AdS_{2n+1}} = - \ft14\big(d\tau + e^\phi(d\chi- x_i\, dy_i)\big)^2
+\ft14 d\phi^2 + \ft14 e^\phi\, (dx_i^2 + dy_i^2) + \ft14
e^{2\phi}\, (d\chi - x_i\, dy_i)^2\ .\label{adsmet2}
\ee
Note that this is an AdS$_{2n+1}$ metric of ``unit radius,'' since it
is defined by $Z^a\, \bar Z^b\, \eta_{ab}= -1$. 

As before, by projecting orthogonally to the orbits of $\del/\del\tau$,
we obtain the metric on $\wtd{CP}^n$, which now takes the simple real form
\be
d\Sigma^2_{2n} = \ft14 d\phi^2 + \ft14 e^\phi\, (dx_i^2 + dy_i^2) + \ft14
e^{2\phi}\, (d\chi - x_i\, dy_i)^2\ .\label{cpnmet2}
\ee

    In fact, the parameterisation of the $\wtd{CP}^n$ metric in terms
of the real coordinates $(\phi,\chi, x_i,y_i)$ coincides precisely
with a coset parameterisation in an ``upper-triangular'' gauge.
To see this, let us define the $(n+1)\times (n+1)$ matrix $E_a{}^b$,
which has zeroes everywhere except for a 1 at row $a$ and column $b$.
Let $H=E_0{}^0 -E_n{}^n$.  Then $H$, together with $E_0{}^n$, $E_0{}^i$ and
$E_i{}^n$, where $1\le i\le n-1$, form the generators of the {\it
solvable Lie algebra} of $SU(n,1)$.  (See, for example,
\cite{f1,f2,f3,f4,lpsk3} for discussions of solvable Lie algebras.)  
In other words, the generators
$E_0{}^i$ and $E_i{}^n$ constitute the entire subset of positive-root
generators that have non-vanishing weights under the Cartan generator
$H$.  (The generator $H$ is the non-compact Cartan generator; all the
others are compact in $SU(n,1)$.)  By exponentiating the solvable Lie
algebra generators, we obtain a gauge-fixed parameterisation of points
in the coset $SU(n,1)/U(n)$.  Let us define the coset representative
\be
\v = e^{\fft12\phi\, H}\, e^{-\im\, \chi\, E_0{}^n}\, \Big(\prod_i
e^{\fft1{\sqrt2} x_i\, (E_0{}^i + E_i{}^n)}\Big)\, 
\Big(\prod_i
e^{-\fft{\im}{\sqrt2} y_i\, (E_0{}^i - E_i{}^n)}\Big)\ ,
\ee
where the terms in the product are ordered such that factors with
larger index values $i$ sit to the right of those with smaller $i$.

    From the commutation relations $[E_a{}^b, E_c{}^d]=\delta^b_c\,
E_a{}^d - \delta _a^d\, E_c{}^b$, it follows that
\be
d\v\, \v^{-1} = \ft12 d\phi\, H + \ft1{\sqrt2}\, e^{\fft12\phi}\,
(dx_i - \im\, dy_i)\, E_0{}^i +  
\ft1{\sqrt2}\, e^{\fft12\phi}\,
(dx_i + \im\, dy_i)\, E_i{}^n - \im\, e^\phi\, (d\chi-x_i\, dy_i)\,
E_0{}^n\ .
\ee
The coset metric can then be written as
\be
ds^2= \ft18 {\rm tr}\, \big( (d\v\, \v^{-1})_\perp \big)^2\ ,\label{coset1}
\ee
where $ (d\v\, \v^{-1})_\perp = d\v\, \v^{-1} + (d\v\,
\v^{-1})^\dagger$.  An alternative way of writing the coset metric is 
\be
ds^2 = \ft1{8} {\rm tr}\, \big( {\cal M}^{-1}\, d{\cal M}\big)^2\ ,
\label{coset2}
\ee
where ${\cal M} = \v^\dagger\, \v$.  It is straightforward to verify
that the metric (\ref{coset1}) or (\ref{coset2}) coincides precisely 
with (\ref{cpnmet2}).  Note that the $SU(n,1)$ symmetry of the metric
can be seen from the Iwasawa decomposition, which implies that there
exists a $U(n)$ compensating transformation ${\cal O}$ such that $\v'= 
{\cal O}\, \v\, \Lambda$ is again in the exponentiation of the
solvable Lie algebra, where $\Lambda$ is an arbitrary constant
$SU(n,1)$ matrix.

\subsection{The boundary of AdS$_{2n+1}$}

    In the AdS/CFT correspondence, the conformal field theory is
defined on the boundary of AdS described in terms of horospherical
coordinates \cite{mald,gkp,wit2}.  To be precise, the horospherical
metric on AdS$_{d+1}$ can be written as
\be
d\Omega^2_{AdS_{d+1}} = d\rho^2 + \eta_{\mu\nu}\, e^{2\rho}\,  
d\xi^\mu\, d\xi^\nu\ ,\label{horo}
\ee
where $\eta_{\mu\nu}$ is the Minkowski metric in $d$ dimensions.  Thus
the boundary at fixed $\rho$ is flat $d$-dimensional Minkowski
spacetime.  In the AdS/CFT correspondence, one considers the boundary 
where $\rho$ tends to infinity.  

    In order to relate our description of AdS$_{2n+1}$ to this
horospherical parameterisation, it is useful first to recall how the metric
(\ref{horo}) on AdS$_{d+1}$ is embedded in flat $\R^{2,d}$.  Introducing real
coordinates $X$, $Y$ and $W^\mu$ in $\R^{2,d}$, we have
\be
X+Y= e^\rho\ ,\qquad X-Y = e^{-\rho} + \xi^\mu\, \xi_\mu\, e^\rho\ ,
\qquad W^\mu = \xi^\mu\, e^\rho\ ,\label{xyw}
\ee
where $-X^2 + Y^2 + W^\mu\, W_\mu = -1$.  Comparing with the complex
coordinates $(Z^0,Z^i,Z^n)$ given in (\ref{xycoord}), which satisfy
the $\R^{2,2n}$ constraint (\ref{const}) (with $d=2n$), 
we see that we can make the identifications 
\bea
Z^0 &=& X + \im\, W^0\ ,\qquad Z^n = Y + \im\, W^1\ ,\nn\\
Z^i &=& W^{2i} + \im\, W^{2i+1}\ ,\qquad 1\le i\le n-1\ .\label{ident}
\eea

    For general values of the AdS$_{2n+1}$ coordinates, this identification
implies that there will be rather complicated relations between the
horospherical description using $(\rho, \xi^\mu)$, and the Hopf-fibred
description using $(\tau,\phi,\chi,x^i,y^i)$.  These can be expressed
as follows:
\bea
&&\tan\ft12\tau = \xi^0+\xi^1\ ,\qquad e^{\fft12\phi} = \sec\ft12\tau\,
e^\rho\ ,\nn\\
&&x^i = 2\cos\ft12\tau\, (\cos\ft12\tau\, \xi^{2i} + 
                          \sin\ft12\tau\, \xi^{2i+1})\ ,\nn\\
&&y^i = 2\cos\ft12\tau\, (\cos\ft12\tau\, \xi^{2i+1} - 
                          \sin\ft12\tau\, \xi^{2i})\ ,\label{rels}\\
&&\chi = \cos^2\ft12\tau\, (\xi^0-\xi^1) + \ft12 x_i\, y_i -
\ft12\sin\tau\, (\xi^\mu\, \xi_\mu + e^{-2\rho})\ .\nn
\eea
In particular, we see that large positive values of $\rho$ correspond to large
positive values of $\phi$, although the $\rho=$constant surfaces do
not coincide with $\phi=$constant surfaces unless $\tau$ is also fixed.  
Note that the time coordinate $\tau$ of the Hopf fibred description
is related to the light-cone coordinate $\xi^+\equiv \xi^0+\xi^1$ of the
horospherical description.

    Let us now consider the boundary metric at fixed $\phi$ in
detail.  From (\ref{adsmet2}), we see that when $\phi$ is large and
is held fixed, the AdS$_{2n+1}$ metric approaches $\ft14 e^\phi\,
ds^2$, where the boundary metric $ds^2$ is given by
\be
ds^2 =  - 2 d\tau\,(d\chi- x_i\, dy_i) + dx_i^2 + dy_i^2 
\ .\label{boundary}
\ee
Unlike the $\rho=$constant boundary in the horospherical description, this
is not a flat metric.  However, it is not hard to show that its Weyl
tensor vanishes, and hence it is conformally flat.  Indeed, if we
define the conformally-rescaled boundary metric $d\td s^2= \Omega^2\,
ds^2$, then we can show that $d\td s^2$ will be flat if $\Omega$ is
taken to be
\be
\Omega = \fft{1}{\cos(\ft12\tau+\alpha)}\ ,
\ee
where $\alpha$ is any constant. This can in fact be understood from the
relation between $\phi$ and $\rho$ given in (\ref{rels}).  

    It is appropriate to make some further remarks about the relation
between various different descriptions of anti-de Sitter spacetime.
In particular, one might wonder how it can be that the time coordinate
$\tau$ in the Hopf-fibred description (\ref{adsmet2}) of the
AdS$_{2n+1}$ metric is periodic, whilst the time coordinate $\xi^0$ in
the horospherical metric (\ref{horo}) is not.  Of course this question
is not a new one that occurs only because of our use of a Hopf
fibration to describe AdS$_{2n+1}$; the same issue arises if one uses
``standard'' coordinates of a traditional kind in AdS, as described,
for example, in \cite{hawk}.  Generalising to $D$ dimensions, we can
write the standard metric on AdS$_D$ as
\be
d\Omega^2_{AdS_{D}} = -\cosh^2 r\, dt^2 + dr^2 + \sinh^2 r\, 
d\Omega^2_{S^{D-2}}\ ,\label{hawkmet}
\ee
where $d\Omega^2_{S^{D-2}}$ denote the metric on a unit
$(D-2)$-sphere.  This can be embedded in a flat $(D+1)$ dimensional
spacetime $\R^{2,D-1}$ with coordinates $(T,S,U^a)$ as follows:
\be
T= \cosh r\, \cos t\ ,\qquad S = \cosh r\, \sin t\ ,\qquad U^a = \sinh
r \, u^a\ ,\label{hawkcor}
\ee
where the $(D-1)$ quanities $u^a$ satisfy $u^a\, u^a=1$, so that the
metric on the unit $(D-2)$-sphere is given by $d\Omega^2_{S^{D-2}}=
du^a\, du^a$ subject to this constraint.  The coodinates of
$\R^{2,D-1}$ thus satisfy $-T^2 -S^2 + U^a\, U^a = -1$.  It is manifest from
(\ref{hawkcor}) that the AdS$_{D}$ metric (\ref{hawkmet}) is given by 
$d\Omega^2_{AdS_{D}} = -dT^2 -dS^2 + dU^a\, dU^a$.  Again we see that
the time coordinate $t$ in the AdS metric is periodic.  One can easily
express the coordinates $(t,r,u^a)$ of the AdS metric (\ref{hawkmet})
in terms of those of the horospherical metric (\ref{horo}) by making
appropriate identifications of the corresponding embedding
coordinates, for example by taking $T=X$ and $S=W^0$, with all the
remaining coordinates $U^a$, which are spacelike, equated with the $Y$
and remaining $W^\mu$ ($\mu>0$) coordinates of the horospherical embedding.

    The global structure of the horospherical metric on AdS was
discussed in detail in \cite{dgt}.  In particular, it was shown that
the horospherical coordinates cover only one half of AdS.  This is
easily seen by noting from (\ref{xyw}) that the positivity of $e^\rho$
implies that only the region $X+Y>0$ of the AdS hyperboloid in
$R^{2,D-1}$ is covered by the horospherical parameterisation.  From
this observation, the dichotomy between the non-periodic time
in the horospherical parameterisation, and the periodic time of the
standard or the Hopf-fibred parameterisations, becomes understandable,
since the latter two descriptions cover the whole of AdS.  In fact the
$\rho=$constant boundary of the horospherical description really
corresponds to the intersection of a 45-degree null hyperplane with
the hyperboloid in $\R^{2,D-1}$, and the time coordinate $\xi^0$
parameterises an infinite arc within the intersection.

   Finally in this section, it is worth commenting on the nature of
the periodic times in the standard parameterisation and the Hopf-fibred
parameterisation of AdS.  In (\ref{hawkmet}), it is evident that
$\del/\del t$ is a Killing vector, and hence in principle one could
perform a timelike reduction on this circle.  However, the result
would be rather inelegant, since the radius of the circle depends on
the radial coordinate $r$ of the AdS metric.  By contrast, in the
timelike Hopf reduction that we consider in this paper, the timelike
Killing vector $\del/\del\tau$ of the AdS metric (\ref{adsmet2}) is of
constant length, and so the reduction circle is of uniform radius over
the entire AdS spacetime.   This gives a much better-behaved
interpretation to the Kaluza-Klein reduction.  It is analogous to the
distinction between reduction on a circle of latitude,
versus a Hopf circle, in a spacelike reduction on a Killing direction
in a sphere \cite{dlp1,dlp2,dlp3}.

\section{Supersymmetry of the AdS$_5\times S^5$ solution}

   In section 2 we showed how the AdS$_5\times S^5$ solution of the
type IIB theory could be dimensionally reduced on the timelike Hopf
fibres of AdS$_5$, viewed as $U(1)$ bundle over $\wtd{CP}^2$.  In
section 3, we showed how the Fubini-Study metric on $\wtd{CP}^2$ could
be constructed in a convenient coordinate system, given by
(\ref{cpnmet2}) with just a single pair of coordinates $x$ and $y$:
\be
d\Sigma_4^2 = \ft14 d\phi^2 + \ft14 e^\phi\, (dx^2 + dy^2) + \ft14
e^{2\phi}\, (d\chi - x\, dy)^2\ .\label{cp2met2}
\ee
The metric on the unit AdS$_5$ is then, from (\ref{adsmet2}), given by
\be
d\Omega_{AdS_5}^2 = - \ft14\big(d\tau + e^\phi(d\chi- x\, dy)\big)^2
+\ft14 d\phi^2 + \ft14 e^\phi\, (dx^2 + dy^2) + \ft14
e^{2\phi}\, (d\chi - x\, dy)^2\ .\label{ads5met}
\ee

     The relevant equation governing the supersymmetry of the
AdS$_5\times S^5$ solution is the Killing spinor equation, which in
terms of the unit AdS$_5\times S^5$ spaces reduces to the conditions
\be
D_\mu\, \ep = \ft12 \Gamma_\mu\, \ep\ ,\qquad
D_m\, \ep = \ft{\im}{2} \Gamma_m\, \ep\ ,
\ee
in AdS$_5$ and $S^5$ respectively.
Note that in writing these equations, we have made a specific choice
of orientation for the solution.  With other choices, the right-hand
sides of these equations could be multiplied by factors of $(-1)$.  

    When we make the Kaluza-Klein reduction to $D=9$, a truncation to
the massless sector involves taking all the fields to be independent
of the reduction coordinate, which in our case will be the time
coordinate.  Consequently, this truncation will project out any
fields, including the Killing spinors, that are time-dependent.  The
relevant Killing spinors of AdS$_5$ are given in equations (\ref{ads5plus})
in the appendix.  We see that there are four of them, and that they
all have specific non-trivial dependence on the time coordinate
$\tau$.  Thus they will all be projected out in the Kaluza-Klein
reduction and truncation process.  Note that the same thing will
happen if the opposite orientation choice for the solution is taken;
the Killing spinors in AdS$_5$ that satisfy the equation $D_\mu\, \ep
= -\ft12\Gamma_\mu\, \ep$ are given in (\ref{ads5minus}), and they too
all have non-trivial time dependence.

    In fact not merely the Killing spinors, but {\it all} spinors are
projected out in the Kaluza-Klein reduction and truncation procedure,
in this Hopf reduction on AdS$_5$.  The reason for this is that the
non-compact $\wtd{CP}^2$ manifold, like the usual compact $CP^2$, does
not admit an ordinary spin structure.   It does allow generalised
spinors, but these must all carry electric charge and be minimally
coupled to the Kaluza-Klein vector potential ${\cal A}_\1$ \cite{hp}.
Such states, with the necessary electric charges, are in fact precisely
non-zero modes in the Kaluza-Klein spectrum.  Thus the full
Kaluza-Klein spectrum of fermionic states in $D=9$ will be associated
with non-zero modes of the massive sectors of the theory.

    The T-duality between the type IIB and type IIA$^*$ theories can
be seen among the massless modes, at the level of the field theories.
Namely, the massless fields from the $S^1$ reduction of type IIB
supergravity map over into massless fields of type IIA$^*$
supergravity, as we discussed in section 2.  The T-duality will also
be visible in the full string spectra, where in $D=9$ we include
massive modes arising both from the Kaluza-Klein towers of states, and
also from the massive string excitations.  The massive Kaluza-Klein
states and string states interchange under T-duality.  Since, as we
have described, the fermions in the $D=9$ type IIB field theory
picture are all contained in the massive Kaluza-Klein sector, it
follows that after T-dualising to the type IIA$^*$ picture, they will
be associated with massive string states.  These will not be seen at
the level of the field theory in the type IIA$^*$ picture, and so from
a field-theoretic standpoint the fermions, and in particular the
supersymmetry, will appear to be ``lost'' in the Hopf dualisation
procedure.  It will, eventually, be regained in the full string
picture.

   The upshot of this is that after dualising to the type IIA$^*$
theory in $D=9$, we have a solution of the form $\wtd{CP}^2\times S^5$
that appears to lack not only supersymmetry, but also any fermions at
all.  Likewise, after lifting back to $D=10$ we will have a type
IIA$^*$ solution of the form $\wtd{CP}^2\times S^1 \times S^5$ without
supersymmetry or fermions.  Nonetheless this solution, from the
bosonic point of view, satisfies all the conditions for being a BPS
state, since it is merely a T-duality transformation of a known BPS
solution, namely the AdS$_5\times S^5$ solution of type IIB.   

\section{Hopf reduction of AdS$_7\times S^4$}

     In \cite{np} it was shown that the AdS$_4\times S^7$ solution of
$D=11$ supergravity could be re-interpreted as an AdS$_4\times CP^3$
solution of the type IIA theory in $D=10$, by performing a dimensional
reduction on the Hopf fibres of $S^7$.  In \cite{dlp1}, this solution was
discussed from the viewpoint of M-theory and the type IIA string.

    Here, we may carry out a somewhat analogous Hopf reduction of the
AdS$_7\times S^4$ solution of the $D=11$ supergravity, where now we
perform the dimensional reduction on the timelike Hopf fibres of
AdS$_7$ viewed as a $U(1)$ bundle over $\wtd{CP}^3$.  

    The AdS$_7\times S^4$ solution is obtained by taking
\bea
ds_{11}^2 &=& ds^2(AdS_7) + ds^2(S^4)\ ,\nn\\
F_\4 &=& 6m\, \ep_\4\ ,
\eea
where $m$ is a constant and $\ep_\4$ is the volume form on the $S^4$
metric $ds^2(S^4)$.  Substituting into the eleven-dimensional
equations of motion, we find that the Ricci tensors for the AdS$_7$
and $S^4$ metrics are given by
\be
R_{\mu\nu} = -6m^2\, g_{\mu\nu}\ ,\qquad R_{mn} = 12m^2\, g_{mn}\ .
\ee
Thus in terms of standard unit metrics $d\Omega^2_{AdS_7}$ and
$d\Omega^2_{S^4}$ on AdS$_7$ and $S^4$, we therefore have
\be
ds_{11}^2 = \fft1{m^2}\, d\Omega^2_{AdS_7} + \fft1{4m^2}\, 
d\Omega^2_{S^4}\ . 
\ee
From the discussion in section 3, we can therefore write this as
\be
ds_{11}^2 = -\fft1{4m^2}\, (d\tau + e^\phi\, (d\chi - x_i\,
dy_i))^2 +\fft1{m^2}\, d\Sigma_6^2 + \fft1{4m^2}\, 
d\Omega^2_{S^4}\ ,
\ee
where $d\Sigma_6^2$ is the metric on $\wtd{CP}^3$, given in
(\ref{cpnmet2}).  

    Comparing with the standard timelike Kaluza-Klein reduction ansatz
from $D=11$ to $D=10$, for which
\bea
d\hat s_{11}^2 &=& - e^{-\fft16\varphi}\, (dt+ \cA_\1)^2 +
e^{\fft43\varphi}\, ds_{10}^2\ ,\nn\\
\hat F_\4 &=& F_\4 + F_\3\wedge (dt+\cA_\1)\ ,
\eea
we see that the AdS$_7\times S^4$ solution can be interpreted as a
solution of the Euclidean-signatured $D=10$ supergravity with fields
given by
\bea
ds_{10}^2 &=& \fft1{m^2}\, d\Sigma_6^2 + \fft1{4m^2}\,
d\Omega^2_{S^4}\ ,\nn\\
F_\4 &=& \fft{3}{8m^3} \, \Omega_\4\ ,\qquad F_\3 = 0 \ ,\nn\\
\cF_\2 &=& \fft{2}{m}\, J\ ,\qquad \varphi=0\ .
\eea
Here, $\Omega_4$ is the volume form of the unit 4-sphere.
The time coordinate $t$ of the eleven-dimensional metric is related to the 
coordinate $\tau$ on the Hopf fibres by $t=\tau/(2m)$.

    The AdS$_7\times S^4$ solution of $D=11$ supergravity has maximal
(\ie $N=4$) supersymmetry from the viewpoint of seven-dimensional
gauged supergravity.  However, the Kaluza-Klein Hopf reduction to
$D=10$ involves discarding all the modes that have dependence on the
time coordinate $t$.  From the results for the Killing spinors on
AdS$_7$ given in the appendix, we see that of the eight Killing
spinors on AdS$_7$ they either all depend upon $\tau$, or else just 2
out the 8 depend upon $\tau$, depending on the sign choice in the
Killing spinor equation.  This means that in the Kaluza-Klein
reduction, the ten-dimensional $\wtd{CP}^3\times S^4$ solution will
have either no supersymmetry, or $N=3$ supersymmetry, from the $D=7$
viewpoint, depending upon the orientation-choice associated with the 
Hopf fibration.  This is analogous to what was seen in the
Hopf reductions of AdS$_4\times S^7$, which gave either $N=0$ or
$N=6$ four-dimensional supersymmetry, depending on the 
orientation \cite{np,dlp1}.

\section*{Acknowledgements} We are grateful to Eugene Cremmer, Bernard Julia
and Hong L\"u for helpful discussions.

\section*{Appendices}
\addcontentsline{toc}{part}{Appendices}

\appendix

\section{Killing Spinors in AdS}

    In order to study the supersymmetry of the solutions obtained by
performing Hopf reductions on the time coordinate in AdS, it is useful
to construct the Killing spinors in the relevant AdS$_{2n+1}$  backgrounds,
viewed as $U(1)$ bundles over $\wtd{CP}^n$.
Thus, in particular, we are interested in the cases of AdS$_5$
and AdS$_7$ in this paper.  

    To begin, we obtain the spin connection for the $\wtd{CP}^n$
metric given in (\ref{cpnmet2}).  To this end, we define a vielbein
basis as follows:
\bea
e^0     & = & \ft12 d\f\ , \nn \\
e^{0'}  & = & \ft12 e^{\f} (d\chi - x_i\, dy_i)\ , \nn \\
e^i     & = & \ft12 e^{\fft12\f} dy^i\ , \la{vielbeins} \\
e^{i'}  & = & \ft12 e^{\fft12\f} dx^i \ . \nn
\eea
(Our notation for the labelling of tangent-space indices is implicitly
defined by these expressions.)
The spin connection $\omega_{ab}$, defined by $de^a=-\omega^a{}_b\,
\wedge e^b$ and $\omega_{ab}=-\omega_{ba}$, is then given by
\bea
&&\w_{00'} = - 2 e^{0'}\, ,\hhs \w_{0i} = -  e^{i}\, ,\hhs
\w_{0i'} = - e^{i'}\, , \nn \\
&&\w_{0'i} = e^{i'}\, ,\hhs \w_{0'i'} = -e^{i}\, ,\hhs
\w_{ij'} = - \d_{ij} \, e^{0'} \ . \la{espcon}
\eea
Note that the 2-form
\be
J = e^0\wedge e^{0'} + e^i \wedge e^{i'} 
\ee
is closed, as can be seen from the fact that locally $J = \ft12 de^{0'}$.
In fact $J$ is the K\"ahler form on $\wtd{CP}^n$.  

   As we saw in section 3, the metric on AdS$_{2n+1}$ can be written
as a $U(1)$ bundle over $\wtd{CP}^n$, as in (\ref{adsmet2}).  If we
introduce the natural vielbein basis for this AdS metric, namely
\be
\hat e^\tau = \ft12 \big(d\tau + e^\phi\, (d\chi - x_i\, dy_i)\big)\ ,
\qquad \hat e^a = e^a\ ,\label{vielbeins2}
\ee
where $a=(0,0',i,i')$ runs over the indices of the
$\wtd{CP}^n$ vielbein $e^a$ defined above, the the spin connection for
AdS$_{2n+1}$ is given by
\bea
&&\hat\w_{0i} = - e^i\, ,\hhs \hat\w_{0i'} = - e^{i'}\,
  ,\hhs \hat\w_{0\tau} = e^{0'}\, ,\hhs \hat\w_{i\tau} = 
  e^{i'}\, ,\nn \\
&&\hat\w_{0'i} =  e^{i'}\, ,\hhs \hat\w_{0'i'} = -  e^{i}\,
  ,\hhs \hat\w_{0'\tau} = - e^0\, ,\hhs \hat\w_{i'\tau} = -
  e^i \, , \la{adsespcon} \\
&&\hat\w_{00'} = \hat e^\tau \, - 2\, e^{0'} \, ,\hhs
  \hat\w_{ij'} =  \d_{ij} \, ( \hat e^\tau \, - \, e^{0'})\,
  \ . \nn
\eea
Note that the potential term $e^\phi\, (d\chi - x_i\, dy_i)$ in the
vielbein $\hat e^\tau$, which is responsible for the topological twist of
the $U(1)$ fibres, is nothing but $2 e^{0'}$, and $\ft12e^{0'}$ is
the potential for the K\"ahler form $J$ on $\wtd{CP}^n$.

     The Killing spinor equation for this AdS metric, which has ``unit
radius,'' is
\be
{\bf D}_M \e^\pm = \pm \, \ft12 \, {\G}_M \, \e^\pm\ , \la{killing}
\ee
where $\e^\pm$ is the Killing spinor, and ${\bf D}_M$ is the covariant
derivative on AdS,
${\bf D}_M \equiv \pa_M + \ft14 \, {{\w}_M}^{AB} \, \G_{AB}$.  Killing
spinors exist for each choice of sign in (\ref{killing}), although in
a specific supergravity solution, having established orientation
conventions, only one particular choice will correspond to unbroken
supersymmetries.  

    From the expressions in (\ref{vielbeins}), (\ref{vielbeins2}) and 
(\ref{adsespcon}) for the vielbeins and spin connection
of AdS$_{2n+1}$, we therefore find that the Killing spinor equation
can be written as
\bea
\pa_\tau \, \e^\pm    & = & \ft14 \, (\pm \, \G_\tau \, - \, \G_{00'} -
                     \G_{ii'}) \, \e^\pm  \la{pat}  \\
\pa_0 \, \e^\pm    & = & \ft14 \, (\pm \, \G_0 \, - \, \G_{0'\tau}) \,
                     \e^\pm  \la{pa0} \\
\pa_{0'} \, \e^\pm & = & \ft14 \, e^{\f} (\G_0 \, \pm \, \oneone) \,
                     (\G_\tau \, + \, \G_{0'}) \, \e^\pm  \la{pa0'} \\
\pa_i \, \e^\pm    & = & \ft14 \, e^{\fft12 \f} \, [(\G_0 \, \pm \,
                     \oneone) \, (\G_i \, - \, e^{\fft12 \f} \,
                     A_i \, (\G_{0'} \, + \, \G_\tau)) \, - \,
                     \G_{i'} \, (\G_{0'} \, + \, \G_\tau)] \, \e^\pm
                     \la{pai} \\
\pa_{i'} \, \e^\pm & = & \ft14 \, e^{\fft12 \f} \, [(\G_0 \, \pm \,
                     \oneone) \, \G_{i'} \, + \G_i \, (\G_{0'} \, +
                     \, \G_\tau)] \, \e^\pm \la{pai'} \ .
\eea

   In order to make explicit computations of the Killing spinors, we need
to choose a representation for the $\G$-matrices.  In terms of the
standard Pauli matrices $\sigma_i$, we may define the $\Gamma$
matrices in Euclidean dimension $D=2n$ as
\bea
\G_1 &=& \s_2 \otimes \s_3 \otimes \s_3 \otimes \cdots
\otimes \s_3 \otimes \s_3\ ,\nn\\
\G_2 &=& \s_1 \otimes \s_3 \otimes \s_3 \otimes \cdots
\otimes \s_3 \otimes \s_3\ ,\nn\\
\G_3 &=& \oneone \otimes \s_2 \otimes \s_3 \otimes \cdots
\otimes \s_3 \otimes \s_3\ ,\nn\\
\G_4 &=& \oneone \otimes \s_1 \otimes \s_3 \otimes \cdots
\otimes \s_3 \otimes \s_3\ ,\label{gammaeven} \\
\G_5 &=& \oneone \otimes \oneone \otimes \s_2 \otimes \cdots
\otimes \s_3 \otimes \s_3\ ,\nn\\
&&\cdots \cdots \ ,\nn\\
\G_{2n-1} &=& \oneone \otimes \oneone \otimes \oneone \cdots
\otimes \oneone \otimes \s_2\ ,\nn\\
\G_{2n} &=& \oneone \otimes \oneone \otimes \oneone \cdots
\otimes \oneone \otimes \s_1\ .\nn
\eea
In an odd dimension $D=2n+1$, with Lorentzian signature, 
we use the above construction for the
Dirac $\G$-matrices of $2n$ dimensions, and take $\G_{2n+1}$ (in the
timelike direction) to be
\be
\Gamma_{2n+1} =i\,\, \s_3 \otimes \s_3 \otimes \s_3 \otimes \cdots
\otimes \s_3 \otimes \s_3\ .\label{gammaodd}
\ee
The one-to-one correspondence between these indices and the ones
we used for our parameterisation of AdS$_{2n+1}$ will be taken to be
\bea
&&\Gamma_{2p-1}\rightarrow \Gamma_{p'}\ ,\qquad
\Gamma_{2p}\rightarrow \Gamma_p\ ,\qquad 1\le p \le n-1\ ,\nn\\
&&\Gamma_{2n-1} \rightarrow \Gamma_{0'}\ ,\qquad
\Gamma_{2n}\rightarrow \Gamma_0\ , \qquad
\Gamma_{2n+1} \rightarrow \Gamma_\tau\ .
\eea

   In terms of these conventions, we find the following results for
the explicit forms of the Killing spinors in AdS$_3$, AdS$_5$ and
AdS$_7$:

\sse{Killing Spinors for AdS$_3$} 

The set of
Killing spinor corresponding to the positive sign in (\ref{killing}) 
is given by
\be
  \e^+_1= \pmatrix{-e^{-\ft{\f}2}-\im \, \c e^{\ft{\f}2} \cr
           e^{-\ft{\f}2}-\im \, \c e^{\ft{\f}2}}\ ,\qquad
  \e^+_2= \pmatrix{ e^{\ft{\f}2} \cr e^{\ft{\f}2}}\ ,\nn
\ee
The other set, corresponding to the negative
sign in (\ref{killing}), is instead given by
\be
\e^-_1= e^{-\ft{\im}{2} \tau} \pmatrix{1 \cr 0}\ ,\qquad
\e^-_2= e^{\ft{\im}{2} \tau} \pmatrix{0 \cr 1}\ .
\ee

\sse{Killing Spinors for AdS$_5$}

The set of
Killing spinor corresponding to the positive sign in (\ref{killing}) 
is given by
\bea
&&\e^+_1= e^{-\ft{\im}{4} \tau }\pmatrix{\ft12 e^{\ft12 \f}
(y -\im x) \cr  \ft12 e^{\ft12 \f} (y -\im x) \cr
1 \cr 0}\ ,\nn \\[.3cm]
&&\e^+_3 = e^{-\ft{\im}{4} \tau } \pmatrix{-e^{-\ft12 \f} +
e^{\ft12 \f} [-\im \c + \ft12\im y x -\ft14 (y^2+x^2)]\cr
e^{-\ft12 \f} + e^{\ft12 \f} [-\im \c + \ft12\im y x -
\ft14 (y^2+x^2)]\cr- (y+\im x)\cr 0}\ ,\label{ads5plus} \\[.3cm]
&&\e^+_2 = e^{-\ft{\im}{4} \tau }\pmatrix{
e^{\ft12 \f} \cr e^{\ft12 \f} \cr 0 \cr 0}\ ,\qquad
\e^+_4 = e^{\ft{3\im}{4} \tau }\pmatrix{0\cr 0\cr 0\cr 1}\ . \nn
\eea

The other set of Killing spinors corresponding to the negative
sign is instead given by
\bea
&&\e^-_1= e^{\ft{\im}{4} \tau }\pmatrix{0\cr 1\cr - \ft12 e^{\ft12 \f}
(y +\im x) \cr  \ft12 e^{\ft12 \f} (y + \im x)}\ ,\nn \\[.3cm]
&&\e^-_3 = e^{ \ft{\im}{4} \tau } \pmatrix{0\cr (y-\im x)\cr
e^{-\ft12 \f} +
e^{\ft12 \f} [\im \c + \ft12\im y x -\ft14 (y^2+x^2)]\cr
e^{-\ft12 \f} + e^{\ft12 \f} [-\im \c + \ft12\im y x +
\ft14 (y^2+x^2)]}\ ,\label{ads5minus} \\[.3cm]
&&\e^-_2 = e^{\ft{\im}{4} \tau }\pmatrix{ 0\cr 0\cr
- e^{\ft12 \f} \cr e^{\ft12 \f}}\ ,\qquad
\e^-_4 = e^{-\ft{3\im}{4} \tau }\pmatrix{1\cr 0\cr 0\cr 0}\ . \nn
\eea

\sse{Killing Spinors for AdS$_7$}

We find that the set of Killing spinor
corresponding to the positive sign in (\ref{killing}) is given by\\[.3cm]
\bea
&&\hhs \e^+_1= e^{-\ft{\im}{2} \tau}\pmatrix{\ft12 e^{\ft12 \f}
( y_1-\im x_1) \cr  \ft12 e^{\ft12 \f} ( y_1-\im x_1) \cr
0 \cr 0\cr 1 \cr 0 \cr 0 \cr 0}\ ,\qquad
\e^+_2= e^{-\ft{\im}{2} \tau}\pmatrix{\ft12 e^{\ft12 \f}
( y_2-\im x_2) \cr  \ft12 e^{\ft12 \f} ( y_2-\im x_2) \cr
1 \cr 0\cr 0 \cr 0 \cr 0 \cr 0}\ ,\nn \\[.5cm]
&&\e^+_3 = e^{-\ft{\im}{2} \tau} \pmatrix{-e^{-\ft12 \f} +
e^{\ft12 \f} [-\im \c + \ft12\im ( y_1 x_1+ y_2 x_2) -
\ft14 ( y_1^2+ y_2^2+x_1^2+x_2^2)]\cr
e^{-\ft12 \f} + e^{\ft12 \f} [-\im \c + \ft12\im ( y_1 x_1
+ y_2 x_2) - \ft14 ( y_1^2+ y_2^2+x_1^2+x_2^2)]\cr
- ( y_2+\im x_2)\cr 0 \cr - ( y_1+\im x_1)\cr 0 \cr
0 \cr  0}\ , \nn \\[.5cm]
&&\hhs \hhs \hhs \e^+_4= e^{-\ft{\im}{2} \tau}\pmatrix{
e^{\ft12 \f} \cr e^{\ft12 \f} \cr 0 \cr 0 \cr 0 \cr 0
\cr 0 \cr 0}\ ,\qquad \hhs \hhs
\e^+_6= e^{\ft{\im}{2} \tau}\pmatrix{0 \cr 0 \cr 0 \cr
0 \cr 0 \cr 0 \cr e^{\ft12 \f} \cr e^{\ft12 \f}}\ ,
\la{ads7+}\\[.5cm]
&&\hhs \e^+_7= e^{\ft{\im}{2} \tau}\pmatrix{0 \cr 0\cr 0 \cr 0 \cr
0 \cr 1 \cr -\ft12 e^{\ft12 \f} ( y_2+\im x_2) \cr - \ft12
e^{\ft12 \f} ( y_2+\im x_2)}\ ,\qquad
\e^+_8= e^{\ft{\im}{2} \tau}\pmatrix{0 \cr 0\cr 0 \cr 1 \cr 0
\cr 0 \cr \ft12 e^{\ft12 \f} ( y_1+\im x_1) \cr  \ft12
e^{\ft12 \f} ( y_1+\im x_1)}\ ,\nn \\[.5cm]
&&\e^+_5 = e^{\ft{\im}{2} \tau} \pmatrix{0 \cr 0 \cr 0 \cr
( y_1-\im x_1) \cr 0 \cr - ( y_2-\im x_2) \cr -e^{-\ft12 \f} +
e^{\ft12 \f} [-\im \c + \ft12\im ( y_1 x_1+ y_2 x_2) +
\ft14 ( y_1^2+ y_2^2+x_1^2+x_2^2)]\cr
e^{-\ft12 \f} + e^{\ft12 \f} [-\im \c + \ft12\im ( y_1 x_1
+ y_2 x_2) + \ft14 ( y_1^2+ y_2^2+x_1^2+x_2^2)]}\ . \nn
\eea
\\

    The other set, corresponding to the negative
sign in (\ref{killing}), is instead given by\\[.3cm]
\bea
&&\e^-_1= e^{\im \tau}\pmatrix{0 \cr 0 \cr 0 \cr
0 \cr 0 \cr 0 \cr 0 \cr 1}\ ,\hs
\e^-_8= e^{- \im \tau}\pmatrix{1 \cr 0 \cr 0 \cr
0 \cr 0 \cr 0 \cr 0 \cr 0}\ ,\qquad
\e^-_3 =\pmatrix{0 \cr 0 \cr 0 \cr 0 \cr - e^{\ft12 \f} \cr
e^{\ft12 \f} \cr 0 \cr 0}\ ,\qquad
\e^-_5 =\pmatrix{0 \cr 0 \cr - e^{\ft12 \f} \cr
e^{\ft12 \f} \cr 0 \cr 0 \cr 0 \cr 0}\ ,\nn \\[.5cm]
&&\e^-_4 =\pmatrix{0 \cr ( y_1-\im x_1) \cr -\ft12
e^{\ft12 \f} ( y_1-\im x_1) ( y_2+\im x_2) \cr
\ft12 e^{\ft12 \f} ( y_1-\im x_1) ( y_2+\im x_2) \cr
e^{-\ft12 \f} - e^{\ft12 \f} [-\im \c + \ft12\im ( y_1 x_1+
 y_2 x_2) + \ft14 ( y_1^2- y_2^2+x_1^2-x_2^2)]\cr
e^{-\ft12 \f} + e^{\ft12 \f} [-\im \c + \ft12\im ( y_1 x_1+
 y_2 x_2) + \ft14 ( y_1^2- y_2^2+x_1^2-x_2^2)]\cr
- ( y_2+\im x_2) \cr 0}\ , \nn \\[.5cm]
&&\hhs \e^-_2 =\pmatrix{0 \cr 1 \cr  -\ft12 e^{\ft12 \f}
( y_2+\im x_2) \cr \ft12 e^{\ft12 \f} ( y_2+\im x_2) \cr
- \ft12 e^{\ft12 \f} ( y_1+\im x_1) \cr \ft12 e^{\ft12 \f}
( y_1+\im x_1) \cr 0 \cr 0}\ ,\qquad
\e^-_7 =\pmatrix{0 \cr 0 \cr \ft12 e^{\ft12 \f}
( y_1-\im x_1) \cr - \ft12 e^{\ft12 \f} ( y_1-\im x_1) \cr
-\ft12 e^{\ft12 \f} ( y_2-\im x_2) \cr  \ft12 e^{\ft12 \f}
( y_2-\im x_2) \cr 1 \cr 0}\ ,\la{ads7-} \\[.5cm]
&&\e^-_6 =\pmatrix{0 \cr ( y_2-\im x_2) \cr e^{-\ft12 \f} -
e^{\ft12 \f} [-\im \c + \ft12\im ( y_1 x_1+ y_2 x_2) + \ft14
( y_2^2- y_1^2+x_2^2-x_1^2)]\cr e^{-\ft12 \f} + e^{\ft12 \f}
[-\im \c + \ft12\im ( y_1 x_1+ y_2 x_2) + \ft14 ( y_2^2- y_1^2+
x_2^2-x_1^2)]\cr -\ft12 e^{\ft12 \f} ( y_1+\im x_1)
( y_2-\im x_2) \cr \ft12 e^{\ft12 \f} ( y_1+\im x_1)
( y_2-\im x_2) \cr ( y_1+\im x_1) \cr 0}\ . \nn
\eea

\end{document}